\documentclass[prl,twocolumn,showpacs,preprintnumbers,amsmath,amssymb]{revtex4}
\usepackage{graphicx}
\usepackage{dcolumn}
\usepackage{bm}

\newcommand{\Tc}{T_{\mathrm{C}}}

\begin{document}

\title{Thermal collapse of spin-polarization in half-metallic ferromagnets}
        
\author{M.~Le\v{z}ai\'{c}}\email{M.Lezaic@fz-juelich.de}
\author{Ph.~Mavropoulos}\email{Ph.Mavropoulos@fz-juelich.de}
\author{J.~Enkovaara}
\author{G.~Bihlmayer}
\author{S.~Bl\"ugel}

\affiliation{Institut f\"ur Festk\"orperforschung, Forschungszentrum
  J\"ulich, D-52425 J\"ulich, Germany}


\begin{abstract}
The temperature dependence of the magnetization and spin-polarization
at the Fermi level is investigated for half-metallic ferromagnets. We
reveal a new mechanism, where the hybridization of states forming
the half-metallic gap depends on thermal spin fluctuations and the
polarization can drop abruptly at temperatures much lower than the
Curie point.  We verify this for NiMnSb by {\it ab initio}
calculations. The thermal properties are studied by mapping {\it
ab initio} results to an extended Heisenberg model which includes
longitudinal fluctuations and is solved by a Monte Carlo method.
\end{abstract}


\maketitle

Half-metallic ferromagnets (HMFs) are ferromagnetic metallic compounds
showing, in the ideal case and at zero temperature, a spin
polarization of $P=100\%$ at the Fermi level $E_F$. This means that
the spin-resolved density of states (DOS) shows a metallic character
only for one spin direction (usually majority-spin), with energy bands
crossing $E_F$; contrary to this, the other spin direction
(minority-spin) behaves like an insulator with $E_F$ inside a band gap.
This exotic behavior has inspired research not only in the
field of basic science, but also for applications in spintronics,
since the extreme spin polarization suggests that HMFs are ideal for
inducing and manipulating the transport of spin-polarized electrons.

Since HMFs were originally introduced in 1983~\cite{deGroot83}, their
properties have been explored extensively. Theoretical studies have
focused mainly on their ground-state properties: the magnetic moments
and the origin of the gap~\cite{Galanakis02a}. The stability of the
gap was studied with respect to spin-orbit
coupling~\cite{Mavropoulos04}, to surface and interface
states~\cite{Jenkins01}, to the presence of defects~\cite{Picozzi04}
and to the appearance of non-quasiparticle states~\cite{Chioncel}.  In
parallel, the extreme spin polarization has been verified
experimentally in a few compounds~\cite{Hanssen90}. However, it is
clear that the ideal half-metallic property ($P=100\%$) cannot be
present at elevated temperatures. Fluctuations of magnetic moments
will mix the two spin channels, and at latest at the Curie point,
$\Tc$, the spin polarization will vanish together with the
magnetization. Thus, for application purposes one seeks HMFs with
$\Tc$ significantly higher than room temperature, reasonably assuming
that the temperature dependence of the spin polarization, $P(T)$,
approximately follows the magnetization, i.e., $P(T)\propto
M(T)$~\cite{Skomski02}. But the theoretical study of $P(T)$ is far
from trivial. Materials specific, first-principles calculations based
on density-functional theory (DFT), which capture the physics of
hybridizations and bonding essential to the half-metallic
property~\cite{Galanakis02a}, are designed in principle for the ground
state and not for excited state properties. Nevertheless, adiabatic
spin dynamics can be approximated within DFT, with successful
applications in the prediction of $\Tc$~\cite{Halilov05,Gyorffy85},
lately also for Heusler alloys~\cite{Sasioglu05,Kurtulus05,Rusz}.

In many cases, HMFs have more than one magnetic atom per unit
cell. For example, NiMnSb, a half-Heusler compound and prototypical
example of all HMF has two magnetic atoms: Mn and Ni.  It is the
$d$-$d$ hybridization between the Mn and Ni minority states that opens
the half-metallic gap~\cite{Galanakis02a}, at least at $T=0$.  At
higher temperatures, directional fluctuations of local moments reduce
the magnetization.  In one-component systems it is well-known that the
magnetic configuration at each instant shows some degree of
short-range order: small regions present almost collinear magnetic
moments, with a local spin quantization axis $\hat{e}_\mathrm{loc}$
not necessarily parallel to the average moment (the global axis
$\hat{e}_\mathrm{glob}$), while the low-energy, long-wavelength
fluctuations are more significant for the decrease of the
magnetization. This behavior continues up to and even above
$\Tc$. However, rather little is known for multi-component systems,
when the sublattices are coupled with different strength so that
they can lose the magnetic order at different temperatures.

Motivated by these considerations, we follow two approaches to the
excited states, both based on DFT (one in connection to an extended
Heisenberg model, and one within mean-field theory), in order to
elucidate the problem of magnetization and spin polarization at
$T>0$. Our focus is on NiMnSb.  We conclude that the hypothesis
$P(T)\propto M(T)$ is not valid. Instead, $P(T)$ can fall off much
faster than $M(T)$.  Furthermore, for NiMnSb we find that the Ni
moment disorders already at very low temperatures, resulting in a
susceptibility peak at 50~K.

\begin{figure}
\includegraphics[angle=270,width=0.99\linewidth]{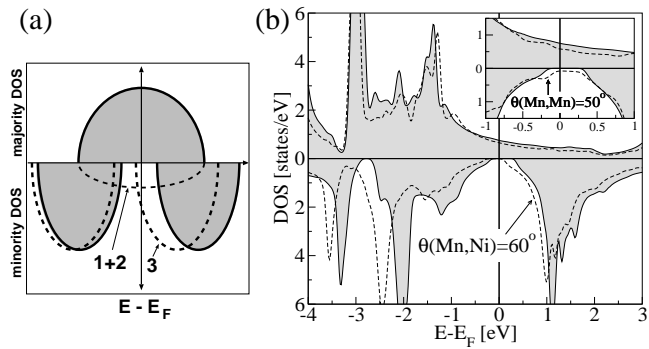}
\caption{(a): Modification of spin-down DOS around $E_F$ reducing
  the spin polarization due to mechanisms (1)--(3) as discussed in
  text.  (b): DOS of NiMnSb in the ground state (full curve) and in
  the state when the Ni moment has been constrained to an angle of
  $60^\circ$ with respect to the Mn moment (dashed curve). Inset:
  Dashed curve: the state when Mn moments of neighboring layers
  along [001] direction form an angle of $50^\circ$ with respect to
  each other, while the Ni moments are allowed to relax.  DOS are
  projected to the global frame. \label{fig:mechanisms}}
\end{figure}

We expect that the decrease of $P$ in the presence of non-collinear
fluctuations arises due to three mechanisms: (1) Firstly, we have a
globally non-collinear effect: the local axis $\hat{e}_\mathrm{loc}$
of a region with short-range order is in general not parallel
$\hat{e}_\mathrm{glob}$. Thus, there is always a projection of locally
spin-up states to the globally spin-down direction in the gap
region. (2) Secondly, a locally non-collinear effect: the short-range
order is not perfect, since the spin axis of each atom varies with
respect to that of its neighbors. This effect can be more significant
in multi-component systems. It is most important if, e.g., a spin-up
localized $d$ resonance determining the local frame is at $E_F$.  The
result of this effect is that, even in a local spin frame of
reference, spin-up wavefunctions of each atom are partly projected
into the spin-down states of its neighbors within the gap, so that $P$
is diminished.  In the presence of only the first effect, $P(T)$ is
expected to follow the average magnetization $M(T)$; the second effect
arises in addition to the first, and $P(T)$ is even further
reduced. (3) Importantly, there is a hybridization effect (change in
hybridization strength), which has been overlooked so-far, leading to
a closure of the gap by shifting the conduction and valence band
edges.  In Fig.~\ref{fig:mechanisms}a we show schematically how the
three mechanisms affect the minority-spin DOS. Naturally, changes in
the majority DOS also occur and charge neutrality is conserved.

We turn now to the specific system, namely NiMnSb. We find by
first-principles calculations that, in the ground state, the Mn local
magnetic moment is large ($M_{\mathrm{Mn}}=3.7\ \mu_B$),
while the Ni local moment is much smaller ($M_{\mathrm{Ni}}=0.26\
\mu_B$) and originates mainly from the hybridization of the $d$
orbitals of Mn with the ones of Ni (the same mechanism which opens the
half-metallic gap~\cite{Galanakis02a}).  In particular, the $d$-$d$
hybridization causes some transfer of weight from Mn to Ni in the
unoccupied states and vice-versa in the occupied ones. Thus, the Ni
moment is formed, while the Mn moment is reduced. Consequently, the Ni
moment does not really arise from strong intra-atomic exchange
interactions (Hund's rule), and its formation lies at a much lower
energy scale than that of the Mn moment. We verify this conclusion
by performing constrained DFT~\cite{Kurz04} 
calculations: when the magnetic moment of Mn is constrained by a
magnetic field to form an angle to its Mn neighbors, it changes very
little, even in an artificial antiferromagnetic
configuration. Contrary to this, the Ni moment vanishes already when
it is constrained at an angle of $90^\circ$ with respect to the Mn
moment, or when the Mn atoms are placed in an artificial
antiferromagnetic order.

We approach the finite-temperature properties starting with a standard
recipe: the adiabatic approximation for the calculation of magnon
spectra~\cite{footnote}.
We employ the full-potential linearized augmented
plane-wave method (FLAPW) as implemented in the {\tt FLEUR}
code~\cite{FLEUR} within the generalized gradient
approximation~\cite{GGA} to DFT.
Total-energy calculations of frozen magnons, performed at the
equilibrium lattice constant of 5.9~\AA, on a dense mesh of 2745
wavevectors $\vec{q}$ and 4096 $\vec{k}$-points, provide the magnon
dispersion relations $E(\vec{q})$. A subsequent Fourier transform
yields the real-space exchange constants $J_{ij}$ between sites $i$
and $j$ ($\in$ \{Ni,Mn\}) \cite{Halilov05}, mapping the system to
a classical Heisenberg model.  Thermodynamic quantities such as the
magnetization curve $M(T)$, susceptibility $\chi(T)$, and Curie point
$\Tc$ can be found within this model by a Monte Carlo method
\cite{Landau}. In multi-component systems we also consider the
sublattice susceptibility $\chi_n$ and magnetization $M_n$ ($n$ is a
sublattice index). A peak of $\chi_n(T)$ signals the release of the
corresponding degrees of freedom, i.e., the sublattice magnetization
is randomized to a great extent; the total susceptibility $\chi(T)$
presents a peak at $\Tc$.

While the Mn moment can be treated within the Heisenberg model as
having a fixed absolute value and fluctuating only in its direction,
for the weak Ni moment the longitudinal fluctuations are energetically
as relevant as the transversal ones. These considerations require an
extension of the traditional Heisenberg model allowing fluctuations of
the magnitude of $M_{\mathrm{Ni}}$.  This is possible, since the
energy needed for constraining the Ni moment can be calculated within
DFT and fitted well by a fourth-order function. Within this
approximation, at each Ni site $i$ the neighboring atoms, placed at
sites $j$, act as an exchange field $\vec{B}_i=\sum_j J_{ij}
\vec{M}_j$ polarizing the Ni atom. Thus, the energy expression for the
Ni atom at $i$ includes the magnitude of the \emph{local} moment $M_i$
and the neighbor-induced polarizing field, and the total Hamiltonian
reads
\begin{equation}
H\!=\!-\tfrac{1}{2}\sum_{i,j}J_{ij}\vec{M}_i\cdot \vec{M}_j +\!\!
\sum_{i\in \{\mathrm{Ni}\}}\!\!(a\,M_i^2+b\,M_i^4-\vec{B}_i\cdot\vec{M_i}).
\label{eq:2}
\end{equation}
The constants $a=18.4$~mRy$/\mu_B^2$ and $b=42.6$~mRy$/\mu_B^4$, which
are fitted to the {\it ab initio} total energy results, are both
positive, giving an energy minimum at $M_i=0$ if the neighbors have a
zero net contribution, i.e., if $\vec{B}_i=0$. The second sum of
Eq.~(\ref{eq:2}) has to be applied for all Ni sites $i$ entering the
Monte Carlo calculation, on top of the usual first part, which is just
the Heisenberg expression. Note that, although the Ni moment is small
and resulting from hybridization, a remnant of intra-atomic exchange
still exists in Ni, giving an enhanced on-site susceptibility. This
assists the local moment formation and is reflected in the values of
$a$ and $b$. Thus, the system can be regarded as an alloy of a
strongly magnetic subsystem (Mn) with a paramagnetic subsystem with
Stoner-enhanced susceptibility (Ni).

\begin{figure}
\begin{center}
\includegraphics[angle=270,width=0.99\linewidth]{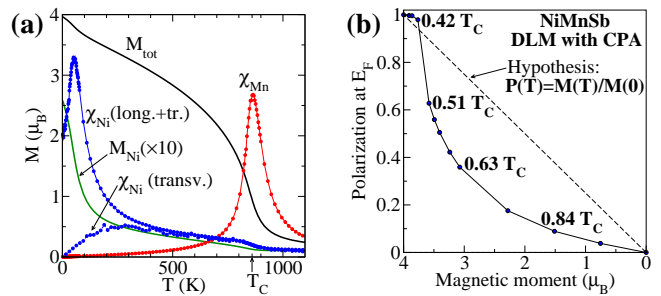}
\caption{(color online) (a): Monte Carlo results for the $T$-dependent
  magnetic properties.  The Ni sublattice moment $M_{\mathrm{Ni}}$
  (magnified by a factor 10) drops fast at 50~K, fluctuating
  transversely and longitudinally; a susceptibility peak
  ($\chi_{\mathrm{Ni}}$(long.+tr.)) is evident. If only transverse
  fluctuations are allowed, $\chi_{\mathrm{Ni}}$(transv.) has a less
  pronounced maximum.  (b): Polarization at $E_F$ as function of total spin
  moment for the DLM states for
  NiMn$_{1-x}^\uparrow$Mn$_{x}^\downarrow$Sb calculated with the
  CPA. The degree $x$ of disorder increases from left ($x=0$\%) to
  right ($x=50$\%), simulating the increase of $T$ from $0$ to $\Tc$.
  \label{fig:2}}
\end{center}
\end{figure}

After calculating the exchange parameters $J_{ij}$ (our results agree
with those of Ref.~\cite{Sasioglu05}), the Monte Carlo calculation
according to Eq.~(\ref{eq:2}) yields the sublattice magnetizations
$M_{\mathrm{Mn}}(T)$ and $M_{\mathrm{Ni}}(T)$, and the susceptibilities
$\chi_{\mathrm{Mn}}(T)$ and $\chi_{\mathrm{Ni}}(T)$, shown in
Fig.~\ref{fig:2}a. Evidently, the overall thermodynamics are governed
by the mighty Mn moment.  The phase transition is clearly seen by the
peak in the susceptibility $\chi(T)$, which grossly coincides with the
Mn sublattice susceptibility $\chi_{\mathrm{Mn}}(T)$. A value of
$\Tc\approx 860$~K is deduced, verified also by the method of cumulant
expansion~\cite{Landau},
and lies between the value of 940~K calculated in~\cite{Rusz} and the
experimental value of $\Tc=730$~K. The surprising feature, however, is
the behavior of $M_{\mathrm{Ni}}(T)$ and $\chi_{\mathrm{Ni}}(T)$.
Already at low temperatures, around 50~K, $M_{\mathrm{Ni}}(T)$ shows a
rapid drop and $\chi_{\mathrm{Ni}}(T)$ a corresponding narrow peak.
If we exclude the longitudinal fluctuations and work within the
traditional Heisenberg model, an unpronounced behavior can be seen
(broad maximum in $\chi_{\mathrm{Ni}}^{\mathrm{transv}}(T)$ at around
$T=300$~K in Fig.~\ref{fig:2}a).  These results show that the Ni
sublattice magnetic order is lost to its great extent.  This behavior
is traced back to the comparatively weak exchange constants $J_{ij}$
of the Ni moments to the neighboring atoms, as we found by the {\it ab
initio} calculations.

It is well-known that the classical Heisenberg model cannot capture
the very low-energy spectrum of the quantum Heisenberg model,
therefore the low-$T$ behavior of $M$ is usually not well
reproduced. However, since the longitudinal fluctuations (essential to
our model) on the Ni sublattice allow for a high rate of energy
absorption around the crossover temperature (50~K), the classical
treatment of the Ni sublattice is applicable already at such low $T$. We
believe the peak of $\chi_{\mathrm{Ni}}$ and the drop of
$M_{\mathrm{Ni}}$ at low $T$ to be connected to the so-far unexplained
experimental findings of an anomaly in the temperature dependence of
the magnetization and the resistivity at approximately
80~K~\cite{Hordequin96}. Assuming that the magnetic moments on the Ni
sublattice are disordered, the system can absorb energy at a higher
rate and this would lead to a higher dissipation and subsequent
increase in resistance.

The loss of short-range order in the Ni sublattice at such low
temperatures suggests, at first sight, that half-metallicity is easily
lost, due to the mechanism (2) of local non-collinearity.  However,
this is not completely correct. Since the gap originates from the
$d$-$d$ hybridization between Ni and Mn spin-down states, and since
the Ni $d$ states are below $E_F$ for both spin directions, a rotation
of the Ni moment causes a hybridization of the Mn $d$ spin-down states
partly with the Ni spin-up states and partly with the Ni spin-down
states. Consequently, the $d$-$d$ hybridization remains, but the gap
width is reduced. The relevant DOS is shown in
Fig.~\ref{fig:mechanisms}b, for the case of the Mn moments remaining
in a ferromagnetic configuration, while the Ni moments are constrained
to a $60^\circ$ angle relative to Mn (and as a result their magnitude
is reduced to 0.16~$\mu_B$).  The reduction of the gap width follows
from a reduction of the hybridization strength: as the Ni moment is
reduced, its spin-down states move lower in energy and hybridize less
with the Mn spin-down states. Thus, the spin-down ``conduction band''
ultimately reaches $E_F$ and half-metallicity is lost. This is the
signature of the hybridization mechanism (3), and comes on top of the
non-collinear behavior of the Mn atoms (mechanisms (1) and (2)). The
latter should be weak at such low $T$, since the average Mn moment
should still be high, dictated by Bloch's $T^{3/2}$ law.  At higher
temperatures the locally non-collinear mechanism (2) should appear for
the Mn moments. In the inset of Fig.~\ref{fig:mechanisms}b we show the
state when Mn moments of neighboring layers form an angle of
$50^\circ$ with respect to each other. Although mechanism (2) produces
a mild effect of a finite spin-down DOS at $E_F$, a protrusion appears
just 0.2~eV under $E_F$. This is once more related to mechanism (3),
triggered by the non-collinear configuration, and can cause a collapse
of $P$ if the protrusion reaches $E_F$.

The results so-far do not contain a quantitative estimate of the
polarization as function of $T$ or of the order parameter $M$.  Such
an estimate requires knowledge of the response of the electronic
structure to the increase of $T$.  We achieve this by proceeding via
mean-field theory, with the disordered local moment (DLM)
state~\cite{Gyorffy85} representing the system at $T>0$ in a
mean-field manner. A Mn site can then have a ``down'' orientation
Mn$_x^\downarrow$ (opposite to the magnetization) with probability
$x$, or an ``up'' orientation Mn$_{1-x}^\uparrow$ (parallel to the
magnetization). The ferromagnetic ground state ($T=0$) corresponds to
$x=0$, while the Curie point corresponds to $x=0.5$. The electronic
structure of the DLM state for each $x$ is found within the coherent
potential approximation (CPA), utilizing the Korringa-Kohn-Rostoker
full-potential Green function method within DFT. This yields the
average magnetic moment $M(x)$ (approximately, $M(x)=M(0)(1-2x)$ so
$x$ is an alternative order parameter), and the polarization $P(x)$.

Considering the above, we study the DLM states,
NiMn$_{1-x}^\uparrow$Mn$_{x}^\downarrow$Sb, with varying concentration
$0<x<0.5$. Within this approach, $P(T)$ cannot be directly found, if
the explicit connection between $x$ and $T$ (or $M(T)$) is not
known. However, one can check the hypothesis $P(T)\sim
M(T)$~\cite{Skomski02}. As shown in Fig.~\ref{fig:2}b, such a
relation does not hold. Instead, from a certain $M$ on, $P$ drops much
faster than $M$. The reason can be traced back to the change in
hybridization as $x$ (or $T$) increases: each Ni atom has on average
$4(1-x)$ Mn$^\uparrow$ neighbors and $4x$ Mn$^\downarrow$
neighbors. The occupied $d$ states of the latter hybridize with the
occupied spin-down states of Ni, push them higher in energy and
diminish the gap; $E_F$ reaches finally the valence band and $P$
collapses. Furthermore, an approximation of $M(T)$ consistent with
mean-field theory can be found by the use of the Brillouin function
$B_j$. Doing this for $j=5/2$, we assigned the
temperature values shown in Fig.~\ref{fig:2}b.  Under this assumption
we see that, up to $T=0.42\ \Tc$ (this is about room temperature),
$P(E_F)$ remains close to 100\%, but then it drops
fast, e.g., $P\approx 35$\% at $T=0.67\ \Tc$.  The globally
non-collinear mechanism (1), not captured by the CPA, should be
present on top of this behavior. Therefore, the initial plateau of $P$
(up to 0.42~$\Tc$) should be corrected towards a linear drop as
$P(T)=M(T)/M(0)$.

In summary, we have investigated the behavior of half-metallic
ferromagnets at elevated temperatures, with emphasis on the properties
of the gap and the spin polarization at $E_F$.  We introduced an
extended Heisenberg model treating transversal as well as longitudinal
magnetic fluctuations, to cover a little investigated situation:
multicomponent magnets which include subsystems of large moments
coexisting with paramagnetic subsystems exhibiting a Stoner-enhanced
susceptibility. The parameters entering the model were determined from
first principles.  We also estimated the polarization at $E_F$ within
a CPA averaging. As prototypical system we have chosen the
half-Heusler compound NiMnSb. Our conclusions are the following: (i)
In NiMnSb, the Ni sublattice is weaker coupled than the Mn one;
longitudinal fluctuations of the Ni moment are energetically as
important as transversal fluctuations. This leads to an early
crossover behavior of the magnetization at $T\approx 0.06~\Tc$ where
the average Ni moment is lost, explaining previous experiments.  (ii)
The hybridization creating the gap is still present but the
fluctuations change its strength. At this stage the gap-width is
reduced. (iii) At higher $T$, fluctuations of the Mn moments introduce
a low DOS into the gap leading to a mild reduction of $P$, and, after
a point, the hybridization changes so much that $E_F$ crosses the band
edges. This is when the polarization collapses (around 0.42~$\Tc$
within mean-field theory).  The behavior of $P$ shown here seems
rather general for half- and full-Heusler alloys exhibiting
half-metallic ferromagnetism as we found by additional
calculations. E.g., for Co$_2$MnSi, the polarization was close to
100\% till 0.27~$\Tc$, then it droped fast, changed sign at 0.63~$\Tc$
and went back to zero at $\Tc$. The decisive factor for the thermal
collapse of polarization is the change in hybridizations due to the
moment fluctuations.  This work calls for experimental efforts to
measure the sublattice magnetization and the spin polarization at
$E_F$ as function of $T$ for half-metals.


We are grateful to P.~H.~Dederichs for enlightening discussions, and
to H.~Ebert and V.~Popescu for providing us with their CPA
algorithm~\cite{SPRKKR}.


\begin{thebibliography}{99}
\bibitem{footnote} Based on the timescale of magnons being much slower
than electron hopping, whence the electron gas has time to relax under
the constraint of the presence of a magnon.
\end{thebibliography}
\end{document}